\documentclass{llncs}
\usepackage{llncsdoc}
\usepackage{graphicx,amssymb,amsmath}
\usepackage{url}
\usepackage{multirow}
\usepackage{color}

\newcommand{\con}[1]{\textsf{\footnotesize #1}}
\newcommand{\ie}{\emph{i.e.}}
\newcommand{\eg}{\emph{e.g.}}
\newcommand{\etc}{\emph{etc.}}

\begin{document}
\title{Hierarchical structuring of Cultural Heritage objects within large aggregations}

\author{Shenghui Wang\inst{1} \and Antoine Isaac\inst{2} \and
  Valentine Charles\inst{2} \and Rob Koopman\inst{1} \and Anthi
Agoropoulou\inst{2} \and Titia van der Werf\inst{1}}

\institute{OCLC Research, Leiden, The Netherlands
\and
Europeana Foundation, The Hague, The Netherlands}

\maketitle
\begin{abstract}
Huge amounts of cultural content have been digitised and are available
through digital libraries and aggregators like Europeana.eu. However,
it is not easy for a user to have an overall picture of what is
available nor to find related objects. We propose a method for
hierarchically structuring cultural objects at different similarity
levels. We describe a fast, scalable clustering algorithm with an automated field selection method for finding semantic
clusters.  We report a qualitative evaluation on the
cluster categories based on records from the UK and a
quantitative one on the results from the complete Europeana
dataset.

\end{abstract}

\section{Introduction}
\label{sec.intro}


More and more Cultural Heritage (CH) content is being digitised and
made available through digital libraries and aggregators such as 
Europeana.eu and the new Digital Public Library of America
(\url{dp.la}). These aggregators provide access to large numbers of
heterogeneous Cultural Heritage objects (CHOs), \eg, Europeana gathers
26 million objects (books, paintings, sound recordings, movies\ldots)
contributed by over 2,200 CH institutions from all over Europe.

Metadata plays a crucial role for these aggregations,
which are largely relying on mappings from the original metadata,
created by providers in many different formats and vocabularies, to a
shared vocabulary like the \textit{Europeana Semantic Elements}
(ESE). However, aggregating metadata from heterogeneous collections
raises quality issues such as uneven granularity of the
descriptions, ambiguity between original and derivative versions of
the same object, 
even duplication if different providers
give access to a same object. Also, simple, common-denominator
vocabularies like ESE, are inappropriate for capturing \emph{internal
  semantic links} between objects (\eg, parts of an object, 
adaptations of a work, objects representing others) 
or \emph{external links} to contextual entities
(
\eg, places or persons related to an
object
).  Both types of
link could benefit services like Europeana by enabling a wider
range of search and browsing options~\cite{Hyvo2012}. 

There are many efforts in the cultural domain to enable and encourage
the provision of richer and interoperable metadata, \eg,
CIDOC-CRM\footnote{\url{http://www.cidoc-crm.org/}} and the new
Europeana Data Model
(EDM).\footnote{\url{http://pro.europeana.eu/edm-documentation}} And
yet, many providers do not have the resources to enhance their
metadata in the way envisioned by these approaches, especially for
links spanning \textit{across} different collections. \textit{Data
  enrichment} in aggregations such as Europeana is therefore valuable.

Meanwhile, keyword-based search is still the main access and
navigation mechanism for such aggregations. Recommendations for
similar object browsing are often provided, \eg, one can ``Explore
further'' in Europeana. Still, in such facilities it is difficult for
a user to have an overall picture of what is available or to find
objects with different levels of relatedness. Researchers have started
looking at automatically identifying related CH
objects~\cite{AlSt2012,Aletras2013,Grieser2011}. However, the existing
work has mostly focused on one dimension of similarity despite the
multidimensional characteristics of the cultural domain. Moreover, it
often stayed at a small scale and could not process datasets as large as Europeana's.

This paper presents a feasibility experiment on semantic linking for a
general, large cultural aggregation. We focus on ``internal''
links between objects from the aggregated collections, with a specific
eye on enabling better-quality ``similar object''
browsing. 
The issue bears
similarity with ``FRBRization'' in the library domain~\cite{oclc-frbr}. However, given
the variety of collections, as well as the simplicity of the
current metadata, it is deemed more realistic to consider a wider
range of object relations: duplication (recognizing records that
describe a same object), depiction/representation, derivation (an
object has been created by reworking another), succession (an object
continues another one), \etc

In this paper, we try to answer the following research questions: (1)
can we apply clustering to find semantic groups at different
similarity levels? (2) what types of useful relationships can we
extract with this technique?

To this end, we propose a framework for hierarchically structuring
objects at different similarity levels in Section~\ref{sec.method},
including a fast and scalable clustering algorithm and an automated
field selection method for finding focal clusters. In
Section~\ref{sec.evaluation}, we report a qualitative evaluation on
the cluster categories based on records from the United Kingdom and a
quantitative evaluation on the results from the complete Europeana
dataset.


\section{Hierarchical structuring based on levels of
similarity}
\label{sec.method}

\begin{figure}[t]
  \centering \includegraphics[width=.8\linewidth]{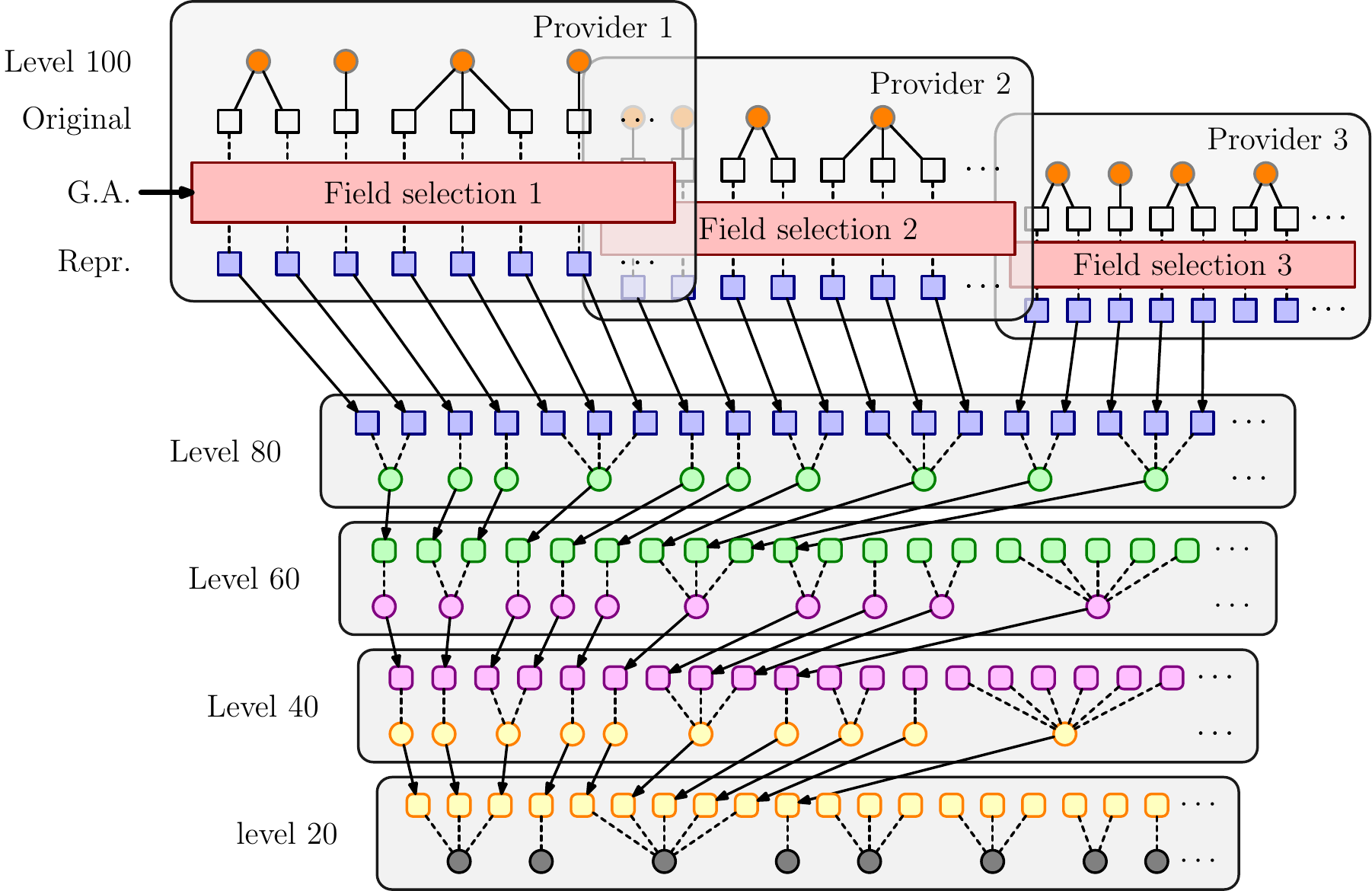}
     \caption{Hierarchical structuring of CHOs at different
       similarity levels. White squares indicate original records
       which are clustered at level 100. Based on genetic metadata
       field selection, the original records are represented by
       selected fields and clustered at level 80. Then clusters at a
       level (circles) are summarised into new artificial records
       (rounded squares at the level below). These are then
       clustered at the lower level, together with the objects that
       were not yet clustered.
       \label{fig.framework}}
  \vspace{-2em}
\end{figure}

We aim at finding related Europeana objects at different levels of
similarity, which potentially reflect different semantic relations
between them. As depicted in Fig.~\ref{fig.framework}, we provide
clusters at five similarity levels. A user can explore the
collections to find CH objects with different levels of relatedness.
We now describe our framework in three parts: (1) fast clustering
based on minhashes and compression similarity
(Section~\ref{sec.clustering}), (2) hierarchically structuring records
at different similarity levels (Section~\ref{sec.hierarchy}) and (3)
automatically selecting important fields based on genetic algorithm to
generate \textit{focal semantic clusters} (Section~\ref{sec.ga})

\subsection{Clustering based on minhashes and compression similarity}
\label{sec.clustering}

\paragraph{Grouping records using combined minhashes}

Records should be clustered based on certain kinds of similarity. 
Because of the sheer amount of
records in the dataset, calculating the pair-wise similarity between
all records is practically impossible and also unnecessary. Therefore
we first group records which could potentially be further clustered
based on a \textit{bag-of-bits} approach.

For each record, the metadata from all fields is combined and divided
into words, with numbers removed. Each word is transformed into
8-shinglings~\cite{MaRaSc2008}. A set of minhashes~\cite{Broder97}
from these shinglings are calculated and randomly put into 4
groups.\footnote{The size of these groups depends on the desired
  similarity level. If clustering at level 100, 16 minhashes are
  randomly chosen for each group, while if at level 20, only 2
  minhashes are selected. In this way, clusters at higher similarity
  levels have higher probability to be precise than those at lower
  levels.} The logical operation exclusive disjunction (XOR) is
applied to each minhash group, producing 4 combined
minhashes. Thus, every Europeana record is represented by four
combined minhashes. Records with the same combined minhashes are
grouped together, as they are the ones that are most likely to be clustered further on. 

\paragraph{Iterative parallel clustering based on compression similarity}

The clustering process is iterative as follows:
\begin{description}
\item [Step 1] Choose a similarity level and set the maximum
  iteration.\footnote{In our experiments, the maximum iteration is set
    at 5.}
\item [Step 2] Group records based on combined minhashes, as
  described above, and put the groups on a stack
\item [Step 3] Get a group of
  records from the stack if the stack is not empty, otherwise, go to Step 7
\item [Step 4] From the group, randomly select up to 10
  records as \textit{cluster heads} that are not closer than the
  required similarity.
\item [Step 5] Assign each record within this group to its closest
  cluster head, which, after all records are assigned, creates
  candidate clusters.
\item [Step 6] For each candidate cluster, if the average similarity
  between the cluster head and the rest of the records is lower than
  the required similarity, put this group of records on the group
  stack to be further divided. Otherwise, this cluster is considered to be a real
  cluster. All the records are considered as \textit{clustered} to the
  cluster head and will not join the next iteration.
\item [Step 7] Collect all the records which are not clustered,
  together with the current cluster heads, repeat Step 3 to 6, until
  no more records can be clustered or the maximum iteration has been reached.
\end{description}

The similarity between records is calculated using a formula adapted
from the Normalised Compression Distance (NCD)~\cite{Cilibrasi05clusteringby}. Let $x$ and $y$ be two
records, $C(xy)$ the compressed size of the concatenation of $x$
and $y$, $C(x)$ and $C(y)$ the compressed size of $x$ and $y$. Then
the similarity between $x$ and $y$ is defined as
\begin{equation*}
  sim(x, y)  = 1.0 - \frac{C(xy) - min(C(x),C(y))}{max(C(x), C(y))}
\end{equation*}

Note, a large part of the clustering process (steps 3 to 6) is
implemented as multi-thread computing, making it very fast and scalable to all
Europeana data.


\subsection{Hierarchically structuring records based on similarity}
\label{sec.hierarchy}

We assume that the clustered records represent a cultural
entity. This is obvious when the similarity level is
high. Clusters at level 100 often contain duplicates (same
object provided twice with the same digital representation) or
near-duplicates (three digitised versions of the same book page). A
cluster at level 80 is often a focal semantic cluster (see
Section~\ref{sec.ga}). These clusters could be clustered at a lower
similarity level too. For instance, pictures of different buildings
are clustered at a lower similarity level, and these pictures could be clustered with census data about the same area at
an even lower level.

Therefore, after each round of clustering at one similarity level, we
generate an artificial record from each cluster, summarising the
information of all the clustered records. These artificial
records, together with all the records which could not be clustered at
this similarity level, will join the clustering process at a lower
similarity level. In this way, hierarchies of records are generated,
so that one can have some structural information about these records,
instead of quickly getting drowned in the sheer amount of data.

In our experiments we activate this hierarchical grouping only below
level 80 because we want to apply a specific field selection process
for this level, which requires all objects.

\subsection{Field selection at level 80 for focal semantic clusters}
\label{sec.ga}

Checking a set of clusters at a high similarity level (\eg, level 80),
one can easily find out that some clusters are of specific interest,
for example, pages of the same book, parts of a same building, \etc
These are often found within a collection from one data provider. They
are more loosely connected than clusters of (near-)duplicates, because they
gather different cultural objects. Yet records from these clusters
collectively represent a small cultural entity.  We name these
clusters \textit{focal semantic clusters} (FSC). These FSCs can be
further clustered at lower similarity levels as described in
Section~\ref{sec.hierarchy}.

However, detecting such FSCs is not easy. The Europeana data is
obtained from a wide range of providers. The information associated
with each record is not uniform, since providers use different
metadata schemes originally and enrich their records with different
amounts of textual information. Take the example of digitised book
pages. One provider may assign exactly the same metadata to all the
pages of the same book while another may give a detailed description
of each page of an illuminated manuscript. For the latter, if all the
metadata fields were used for clustering, the large body of
descriptive texts could falsely separate pages of the same book into
different smaller clusters. It is therefore important to select the
most important metadata fields for clustering these FSCs. As shown in
Fig.~\ref{fig.framework}, such selection is done on a data provider
basis.


For each data provider, we aim at the selection of metadata
fields which gives the best FSCs. We apply a genetic algorithm (GA) to
automatically select important fields, that is, taking an evolutionary
approach to select the optimal solution based on a fitness
function~\cite{Mitchell1999}. This algorithm handles candidate solutions 
as binary sequences, ``1'' when a metadata field is
selected and ``0'' otherwise. For
example, if a given institute provides metadata records with
\con{dc:title}, \con{dc:contributor}, \con{dc:subject} and
\con{dc:source}, then a candidate solution 1010 indicates clustering
on \con{dc:title} and \con{dc:subject} only. In the Europeana
dataset, \con{dc:title} is the most used and often the only
descriptive field. Given its importance, we therefore decided to set it as compulsory for each
data provider's solutions.\footnote{Note, either \con{dc:title} or
  \con{dc:description} are mandatory for data input in Europeana; when
  \con{dc:title} is not available, we take \con{dc:description} as
  the compulsory field.}


The fitness function is to evaluate how good a solution, \ie, a
selection of metadata fields, is to produce reasonable clusters. We
adapted a measure of variance ratio clusterability~\cite{AcBe2009} as
our fitness function: Let X
be a dataset, and $C$ a set of clusters over $X$. The fitness
function is defined as following:
\begin{equation}
  f(C, X)=log(Avg(C)) \times \frac{B_C(X)}{W_C(X)}
\end{equation}
where $B_C(X)$ is the between-cluster distance, $W_C(X)$ is the
within-cluster distances and $Avg(C)$ is the average size of the set
of clusters. This function gives higher fitness (and thus a higher
chance to be selected for the next generation) to tightly connected
clusters that are  relatively big and far apart.

For the genetic evolution, clustering is set at level 80: first
qualitative insights (see Section~\ref{sec.qualitative}) hinted that
this was the ``sensitive level'' for finding such FSCs. The original
records are represented by the metadata from the fields selected in
the GA best solution, and clustered again at level 80.

When clustering at level 60 and lower, other fields
are all taken into account, as this invites broadly linked records
to be clustered and potentially corrects the bias towards links
within individual providers at level 80. Note, at level 100 also, all the
metadata is used to find (near-)duplicates.

\section{Results and evaluation}
\label{sec.evaluation}

\subsection{Qualitative evaluation and categorisation of clusters} 
\label{sec.qualitative}

To guide future evaluation efforts while tuning the method above, we
started a qualitative analysis of intermediate results generated from
1.1M records from UK. The analysis started by looking
at the visual representation and metadata of the clustered records on
the Europeana portal.  
We also browsed the ``hierarchy'' of
clusters produced, giving specific attention to how smaller clusters
combine into bigger clusters and allowing us to find meaningful
clusters for a given (set of) object(s), independently from their
level in the hierarchy.

At that stage, clusters were still sometimes rough and our evaluation
not wide enough for obtaining a clear insight on their respective
representativity. However, this semi-principled analysis offered us
precious insight on the typology of groupings--a typology that looked
both useful and relatively complete, \ie, covering a broad extent of
the relations that EDM covers. 

\noindent \textbf{Same objects/duplicate records}
This is the strongest similarity relationship found in
clusters. Europeana datasets come via different channels: individual
institutions, European projects, thematic portals\ldots It is possible
to receive multiple records for the same object from the same
institution.\footnote{For example, see
  \url{http://www.europeana.eu/resolve/record/09307/2FFD07620AFC6500C005DAC1D0AFCF6A31778A88}
  and
  \url{http://www.europeana.eu/resolve/record/09307/772B1D83F4727C4DEEEF763C300D5315FC1EBEAA}}
A quality control failure during the data ingestion process can let
duplicates be published in the Europeana portal. Clustering allows us
to identify these duplicates with a high degree of accuracy; often the
exact same metadata appears in many fields. 

\begin{figure}[htbp]
  \centering
     \includegraphics[width=.12\linewidth]{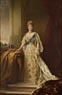}\includegraphics[width=.12\linewidth]{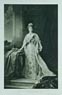}
     \includegraphics[width=.12\linewidth]{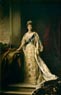}\includegraphics[width=.12\linewidth]{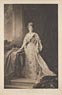}
     \includegraphics[width=.12\linewidth]{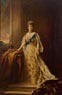}
     \caption{Different views on the same CHO---a portrait of Mary of Teck\label{Views.fig}}
\end{figure}

\noindent \textbf{Views of the same object}
Digitisation practices often lead  providers to create different
views of the same CHO. These views
happen to be provided as different CHOs but they are
actually different views on the same ``real object,'' see
Fig.~\ref{Views.fig}.  Such clusters
usually share exactly the same descriptive metadata. 

\noindent \textbf{Parts of an object}
CHOs provided to Europeana can have a hierarchical structure: they are
composed of other objects or parts. However, digitisation and description choices
by providers, or the barrier of a simplified data format
can
result in the data describing this structure not being provided to
Europeana. 
The clustering
process allows us to find clusters of objects linked by such
relationships. 
In principle relations between different parts of a CHO or between
CHOs should be expressed in relation fields (\con{dc:relation}) but
the clusters indicate that  providers often use \con{dc:title}, see
Table.~\ref{Parts.fig}. In the
latter case, an automatic procedure would have difficulty
making the distinction with other types of relations.
\begin{table}[h]
  \begin{tabular}{|l|c|}\hline
    Shared metadata & Record \\\hline
    \con{dcterms:spatial} : City of London  &  The Oil Shop part 01\\
    \con{dcterms:medium} : Lithograph &The Oil Shop part 03\\
    \con{dc:creator} : Composer: Dallas, John &The Oil Shop part 04\\
    \con{dc:date} : [1873]&The Oil Shop part 05\\
    \con{dcterms:isPartOf} : Victorian popular music. Collect Britain&
    The Oil Shop part 06\\
    \con{dc:format} : jpeg &The Oil Shop part 07\\
    \con{dc:type} : Cover Illustrated Music Printed StillImage &The Oil Shop part 08\\\hline
\end{tabular}
\caption{Parts of a CHO---a music piece made of different individual
  music scores\label{Parts.fig}}
\vspace{-2em}
\end{table}

\begin{figure}[h]
 \vspace{-2em}
 \centering
     \includegraphics[width=.3\linewidth]{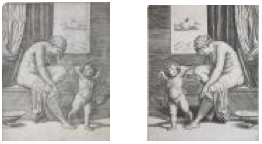}
     \caption{Example of derivative works\label{Derivatives.fig}}
\vspace{-2em}
\end{figure}

\noindent \textbf{Derivative works}
These are objects which are derived from another one, such as
reprint. Fig.~\ref{Derivatives.fig} shows two different prints created
from the same
master.\footnote{\url{http://www.europeana.eu/resolve/record/09405a1/49EADC41C49A4C6F14C626EB067EB7D3F9131632}
  and
  \url{http://www.europeana.eu/resolve/record/09405a1/1A3460CBB5FE76A1CD4433F7FFA052C34A982934}}
Some cases can be analysed in terms of FRBR
relationships, where an original \emph{work}
leads to a range of \emph{expressions}, \emph{manifestations} and/or
\emph{items}. The metadata of the concerned records are often the
same, except the \con{dc:description} field, which usually indicates
that the object is a copy or other type of derivative.

\begin{figure}[h]
\vspace{-2em}
  \centering
     \includegraphics[width=.6\linewidth]{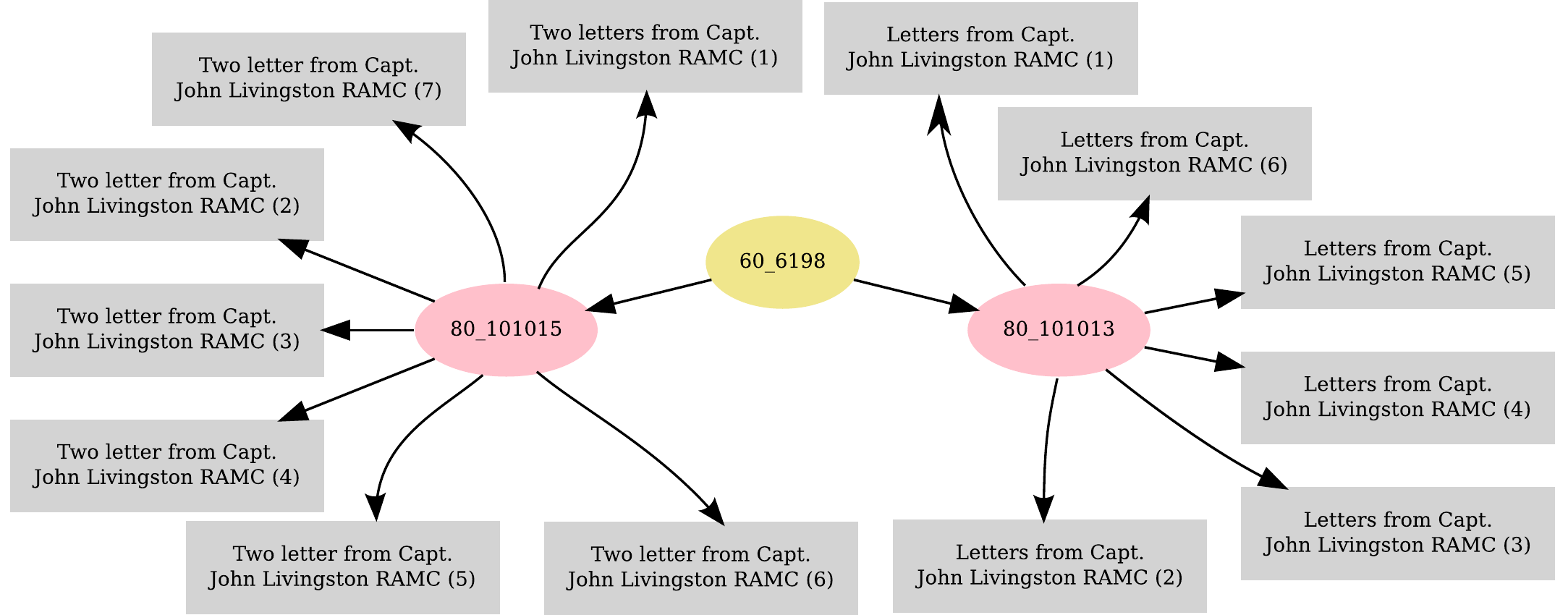}
     \caption{An example of collection clusters\label{Collection.fig}}
\end{figure}

\noindent \textbf{Collections}
Clusters can represent coherent collections.
They group objects of a specific type, gathered by one individual, for
a specific goal. 
For example, the letters shown in
Fig.~\ref{Collection.fig} were written by one specific WWI
soldier and contributed by a family member to the
\emph{Europeana1914-1918} project. Object metadata is often 
similar, with the \con{dc:relation} field expressing
membership in a specific collection.


\noindent \textbf{Thematic groupings}
These clusters gather objects about a similar topic, location or
event, which link them to the other collections above.  However, they
often lack the size or an explicit unity criterion such as common provenance 
(\eg, a collector) that would allow them to be assessed as
complete collections. In fact we have found such individual clusters
included in bigger ones, which have been classified as collections in
the sense above. These clusters have in common some metadata fields
that are related to a similar theme, most often \con{dc:subject}.

\vspace{-1em}
\paragraph{Conclusion} During our qualitative evaluation, we observed that
clusters of ``closely related'' objects, such as duplicates or parts
of a CHO are easier to assess. Recognising clusters describing broader
links, such as topical relationships, seems a more
difficult, error-prone process, both for human evaluators and the
machine. In order to check our finished clustering method, we
proceeded further with a more complete, quantitative evaluation over
the entire dataset.



\subsection{Quantitative manual evaluation on the full
  Europeana dataset}
\label{sec.quantitative}

\textbf{Working dataset}
The entire Europeana data was made available as a dump on
February 2013.  It contains 23,595,555 records from 2428
data providers (defined by \con{europeana:dataProvider} field, or
\con{europeana:provider} when it is not present).

\noindent \textbf{Field selection for FSCs}
\label{sec.gaeval}
1198 individual data providers provided more than 100 records and
cover 99.9\% of the entire dataset.  We applied the genetic
algorithm 
to select the important
fields over these providers. 

We used a python package
Pyevolve,\footnote{\url{http://pyevolve.sourceforge.net/}} setting the
number of individuals at each generation to 50 and the maximum number
of generations to 100. The time taken by field selection depends on the
number of records one provider has. For the 10  providers
with most records, it takes 161 minutes in average, while datasets
with 200-250 records require 21 minutes in average.
Table~\ref{tab.fields} lists the top 5 most selected metadata fields
and the most selected field combinations.  For an overwhelming
majority of data providers, \con{dc:title} carries the most
distinguishing information. In the end, we use the the metadata from
the selected fields for each provider to generate the FSCs at level
80. For the rest of the data providers, we
select \con{dc:title} directly. This leads to 1,476,089 clusters in total. 

\begin{table}[t]
\centering
\begin{tabular}{lcccr}
\begin{tabular}{|c|c|r|}\hline
  & \#Providers &metadata field \\\hline
  1& 2358 &\con{dc:title}\\
  2& 436 & \con{dc:type} \\
  3& 328& \con{dc:language}\\
  4& 315& \con{dc:rights} \\
  5 & 309& \con{dc:subject}\\
  \hline
\end{tabular}

&&&

\begin{tabular}{|c|c|r|}\hline
  &\#Providers &field combination \\\hline

  1& 1521 & \con{dc:title}  \\	
  2& 37 & \con{dc:title} \con{dc:type} \\
  3& 28	& \con{dc:title} \con{dc:creator} \\
  4& 23	& \con{dc:title} \con{dc:identifier} \\
  5& 20	& \con{dc:description} \\ \hline
\end{tabular}
\\
(a) Top 10 most selected fields &&& (b) Top 5 most selected
field combinations\\
\end{tabular}

\caption{Field selection for FSCs (clusters at
  level 80) \label{tab.fields}}
  \vspace{-2em}
\end{table}

\noindent \textbf{Hierarchical structuring of records}
The clustering was carried out on a server with two Intel XEON E5-2670 
processors and 256G memory. Table~\ref{tab.performance} gives the
clustering time per  level. As described in
Section~\ref{sec.hierarchy}, clusters generated at higher levels lead
to artificial records replacing the \textit{clustered} records for the
lower levels. This greatly reduced the amount of items to be clustered
at lower levels.
\begin{table}[h]
  \centering
  \begin{tabular}{|r|r|r|r|}\hline
    Similarity level & \#Records to be clustered & \#Clusters & Time \\\hline
    100 & 23,595,555 & 200,245 & 6m2.82s \\
    80 & 23,595,555 & 1,476,089 & * \\
    60 & 6,407,615 & 382,268 & 3m35.26s\\
    40 & 2,431,753 & 212,389 & 2m28.79s\\
    20 & 1,068,188 & 84,554 & 1m20.99s \\\hline
  \end{tabular}
  \caption{Clustering performance (* Level 80 is clustered
    differently due to the field selection based on GA, see Section~\ref{sec.ga} for more detail.)\label{tab.performance}}
  \vspace{-2em}
\end{table}

\begin{figure}[t]
  \vspace{-1em}
  \centering
     \includegraphics[width=.7\linewidth]{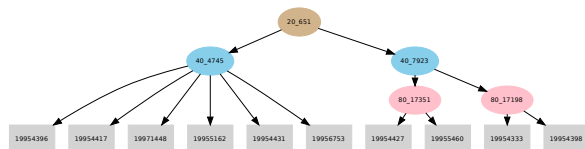}
  \vspace{-2em}
     \caption{Hierarchical view of records
       \label{hierarchy.fig}}
  \vspace{-2em}
\end{figure}

Clusters can be hierarchically ordered across similarity levels.
In Fig.~\ref{hierarchy.fig}, at the record level (the bottom grey
boxes), one can see the sibling records. These can be closely
clustered at level 80 (in pink), or more vaguely connected (at level
40 in blue). These clusters can again be clustered at level 20
(brown). When more records are involved (the size of level 20 clusters
ranges from 2 to 456,155, with an average size of 190), such
structural information is crucial to make sense out of a large amount
of records.

\noindent \textbf{Manual evaluation} To evaluate our method on the
full Europeana dataset and further validate the categories discovered
in Sec.~\ref{sec.qualitative}, we randomly chose 100 clusters at each
level and asked 7 evaluators to categorise them. Clusters were
assigned to evaluators so that each cluster is checked at least by two
 evaluators. For each cluster, the evaluators assigned one of
the six categories from Sec.~\ref{sec.qualitative} or indicated if
it did not make any sense.

The evaluators can leave comments and propose new categories if
necessary. We found that evaluators gave many comments without
proposing any new categories. We measured for each level the average
number of clusters which are assessed as belonging to each category.

\begin{table}[h]
\vspace{-2em}
\centering
  \begin{tabular}{|l| r | r | r | r | r |}\hline
    \multirow{2}{*}{Cluster Category} &
    \multicolumn{5}{|c|}{Similarity Level} \\ \cline{2-6}
    & 100 & 80 & 60 & 40 & 20 \\\hline
    Same objects/duplicate records & 11 & 10 & 1 & 0 & 0\\
    Views of the same object & 61 &33& 6 & 2 & 5\\
    Parts of an object &10 &11& 3 & 1 & 2\\
    Derivative works & 2 &1 & 0 & 0 & 0 \\
    Collections & 1 & 4 &27 & 13 & 43\\
    Thematic grouping & 9 & 34 &36 & 29 & 22  \\
    Nonsense & 2 &3 &30 & 57 & 28 \\\hline
  \end{tabular}
  \caption{Manual evaluation results\label{tab.eval}}
  \vspace{-3em}
\end{table}

As shown in Table~\ref{tab.eval}, duplicate records and views or parts
of the same CHO are mostly clustered at levels 100 and 80. At these
levels we also found different editions of the same work, different
volumes of the same book, pictures of the same event, \etc~Note that
the latter illustrates how thematic groupings also appear as clusters
at the highest level.  Derivative works are rare and only occur at
high levels. Lower levels lead to bigger, more
heterogeneous clusters, many of which fall into the categories of
collections or thematic groupings while many of which do not make much
sense any more. 
These big clusters could contain views of different buildings, issues
of the same journal, different books by the same author, pictures
taken at the same place but at different time, pictures of different
sarcophagi and ships, collections of religious or folk music, thesis
of the same university, specimen of birds, posters about movies or
Communist movements, or more vaguely, collections of furniture or
Spanish books, \etc

The general impression from the evaluators is that most clusters make
sense.  At level 60, it is often clear that the records form a
``collection'' according to some implicit logic; but in most cases the
original provider sites did not present them as explicit
collections. So the clustering was being creative and yet correct.

However, assessing clusters gets
more difficult as the similarity level lowers. It is often difficult
to recognise any specific logic beyond more general and overarching
rules like: ``belonging to same data provider'', ``being of the same
type'', \etc~This is especially so at level 20, where the
average size of the evaluated clusters is 3442, ranging from 2 to
60,204, with 11 clusters having more than 10,000 records.  It is not
possible to manually go through them one by one. Many clusters are
also in a language which the evaluators are not familiar with, which
made them even more difficult to assess. The evaluators only selected
as many sub-clusters as possible to explore the rough structure within
such big clusters.


Of course, not every cluster at level 20 is too big to judge
. For example, one is composed of two (higher-level) sub-clusters that
each corresponds to an edition of a multi-volume book.  While these
are represented as hierarchical objects on the provider's
site\footnote{See
  \url{http://www.biodiversitylibrary.org/bibliography/14916#/summary}
  and \url{
    http://www.biodiversitylibrary.org/bibliography/931#/summary}}
this information could not reach Europeana.  The out-of-the-box SOLR
``MoreLikeThis'' function\footnote{\url{http://wiki.apache.org/solr/MoreLikeThis}} returns the volumes of both editions, but
as a flat, mixed list that includes other books---some configurations tested
for Europeana even fail to bring all volumes of both editions as
related items when one of them is being explored.

In summary, our evaluation shows the clusters are rather
relevant. The two highest levels, especially, could directly
provide meaningful subsets for users of a ``similar items'' browsing
feature.  However, clusters, especially at lower similarity levels,
are much more heterogeneous than we initially thought. We need to make
more detailed distinction between these clusters. The next step of
detecting these different categories automatically is a more
challenging task.

\section{Related work}
\label{sec.relatedwork}

Providing similar objects for access to large collections is not
novel.  Europeana itself uses SOLR's
``MoreLikeThis''.
But such standard search engine features are designed for full-text
documents and suffer from the heterogeneity and sparseness of
the metadata, resulting often in lists that seem random and
unidimensional.  Amazon.com exploit users' input to infuse more
relevance in similar items. But the necessary user data is not
available for cultural aggregators yet. Others have explored using
image similarity instead~\cite{assets} or next to~\cite{AlSt2012}
descriptive metadata. However, digitized content is not available consistently
in cross-domain aggregations, where media types and quality vary greatly.

Tuning textual similarity to CH metadata is therefore still
relevant. \cite{Aletras2013,Knoth2010} have used the standard corpus-based
similarity measures of~\cite{MaRaSc2008}. Recently, researchers started
looking at using external knowledge bases such as
Wikipedia~\cite{Grieser2011} or WordNet~\cite{Mihalcea2006} to help
measuring similarities between objects. Different similarity measures
were compared~\cite{HallCS12,Aletras2013} but most existing work
explore a single dimension of similarity, which does not take into
account the multidimensionality of CH collections; it also focuses on
smaller-scale collections. The extraction of FRBR-like relations, a
topic researched for more than a
decade~\cite{TakhirovTPDL11,oclc-frbr}, has been a clear source for
inspiration for us.  It requires however collections from well-bounded
domains with extensive and consistent metadata, and would need to be
completed with techniques with a broader application scope.  Our work
tries to complement these efforts, further exploring the aspects of
scalability and the typing and organizing of clusters of similar
objects.

\section{Conclusion}
\label{sec.conclusion}
Identifying semantic links and groups of CH objects is desirable
for data enrichment in large cultural aggregations. Finding
similar objects is the first step towards such semantic links. 
Our approach avoids too much dependence on metadata fields and the multidimensionality they denote. Instead, we try to hierarchically structure Europeana objects at different levels, starting with a rather simple similarity measure. We
developed a fast and scalable clustering algorithm and applied a genetic
algorithm to select important fields for generating focal
semantic clusters. We qualitatively evaluated intermediate results
from UK records before carrying out a larger-scale quantitative
evaluation of the results obtained from the entire Europeana dataset.

We found that clusters at higher similarity levels are usually accurate and the
semantic groups make sense to evaluators, \eg, as
duplicates or parts of a CHO. 
The relevance of lower-level clusters is much more difficult
to judge. Even at higher similarity levels, our evaluation shows
that based on a single
dimension of similarity we generate highly heterogeneous clusters. We need to investigate more
multidimensional similarity measures while maintaining the performance levels
for clustering large amounts of data. Future work shall of course include the practical evaluation of hierarchical
structuring for improving end-user navigation.

\bibliographystyle{plain}
\bibliography{clustering_tpdl2013}

\end{document}